\documentclass[aps,preprint,nofootinbib,superscriptaddress]{revtex4}
\usepackage{amssymb}

\begin{document}

\title {Search for possible way of producing super-heavy elements \\
       --Dynamic study on damped reactions of $^{244}$Pu+$^{244}$Pu,\\ $^{238}$U+$^{238}$U
       and $^{197}$Au+$^{197}$Au}

\author{Ning Wang}
\email{wangning@itp.ac.cn} \affiliation{Institute of Theoretical
Physics, Chinese Academic of Science, Beijing 100080, P. R. China}

\author{Zhuxia Li}
\email{lizwux@iris.ciae.ac.cn} \affiliation{China Institute of
Atomic Energy, Beijing 102413, P. R. China} \affiliation{Nuclear
Theory Center of National Laboratory of Heavy Ion Accelerator,
Lanzhou 730000, P. R. China} \affiliation{Institute of Theoretical
Physics, Chinese Academic of Science, Beijing 100080, P. R. China}
\affiliation{Department of Physics, Tsinghua University, Beijing
100084, P. R. China}

\author{Xizhen Wu}
\email{lizwux@iris.ciae.ac.cn} \affiliation{China Institute of
Atomic Energy, Beijing 102413, P. R. China} \affiliation{Nuclear
Theory Center of National Laboratory of Heavy Ion Accelerator,
Lanzhou 730000, P. R. China}

\author{Enguang Zhao}
\affiliation{Institute of Theoretical Physics, Chinese Academic of
Science, Beijing 100080, P. R. China} \affiliation{Department of
Physics, Tsinghua University, Beijing 100084, P. R. China}

\begin{abstract}
  By using the improved Quantum Molecular Dynamics model, the
$^{244}$Pu+$^{244}$Pu, $^{238}$U+$^{238}$U and
$^{197}$Au+$^{197}$Au reactions at the energy range of
$E_{c.m.}$=800 $MeV$ to 2000 $MeV$ are studied. We find that the
production probability of superheavy fragments(SHF) with $Z \geq$
114 is much higher for $^{244}$Pu+$^{244}$Pu reaction compared
with that of $^{238}$U+$^{238}$U reaction and no product of SHF is
found for $^{197}$Au+$^{197}$Au. The production probability of SHF
is narrowly peaked in incident energy dependence. The decay
mechanism of the composite system of projectile and target and the
time scale of decay process are explored. The binding energies of
superheavy fragments are found to be broadly distributed and their
shapes turn out to be exotic form.
\end{abstract} 

\maketitle


  There are two approaches proposed for producing superheavy
  elements (SHE) through accelerators. One approach of the complete fusion reaction
  is very successful in producing SHE. Since the 70's the
elements from $Z$=107 to 116 were synthesized in the "cold fusion"
reactions with lead and bismuth targets \cite{Hof00} and "hot
fusion" reactions with actinide targets \cite{Oga04}. A lot of
research work on this approach have been done both experimentally
and theoretically. However, it is well known that further
experimental extension of the region of SHE to the central of
superheavy "island" with the complete fusion reaction is limited
by the number of available projectiles and targets, and also by
the very low production cross section \cite{Hof00,Oga02}. In order
to explore new more neutron-rich superheavy regions the
radioactive ion beams will have to be utilized, but up to now the
intensive radioactive ion beams are not available. An alternative
pathway to the superheavy elements is the strongly damped
collision process between massive nuclei, for instance $^{238}$U+
$^{238}$U. The strongly damped collisions between
$^{238}$U+$^{238}$U and $^{238}$U+$^{248}$Cm at the energies near
Coulomb barrier were studied in 70s and early 80s for searching
superheavy nuclei \cite{Her79,Sea79,Mar80,Rie79,Gag80}. It was
reported in \cite{Hil77} that for $^{238}$U+$^{238}$U at $E$=7.5
$MeV/nucleon$, the upper limit of the cross-section for producing
superheavy elements was about $2\times10^{-32}$ $cm^{2}$ for half
lives between milliseconds and month by looking for spontaneous
events from reaction products. In \cite{Sch78} the reaction of
$^{238}$U+$^{248}$Cm at 7.4 $MeV/nucleon$ was studied and it was
found that cross sections for $^{100}$Fm, $^{99}$Es, and $^{98}$Cf
with target of $^{248}$Cm are three to four orders of magnitude
higher than with $^{238}$U. It means that the strongly damped
reaction with two nuclei heavier than uranium could be very
benefit for producing superheavy nuclei. Compared with the
approach of complete fusion this approach was much less studied.
Furthermore, considering the fact that  the fast development of
the experimental facilities for searching superheavy nuclei and
deepening understanding of structure properties of superheavy
nuclei in recent years \cite{Rut97,Kru00,Mar01}, the approach of
the strongly damped massive nuclear reaction should be further
studied. In this letter we will study reactions of
$^{244}$Pu+$^{244}$Pu, $^{238}$U+$^{238}$U and
$^{197}$Au+$^{197}$Au at the energy range of $E_{c.m.}$=800
$MeV$-2000 $MeV$ by the microscopically dynamical model. We will
concentrate on 1)the energy-dependence of the probability of
producing the superheavy fragments (SHF) which are defined as the
fragments with charge larger than or equal to 114, 2)the decay
mechanism of the composite system of projectile and target, and
3)the binding energies and shapes of the superheavy fragments.

The Improved Quantum Molecular Dynamics (ImQMD)
Model\cite{Wang02,Hart89,Ai91} is employed in this study. The
application of this model to light and intermediate mass nuclear
fusion reaction was given in \cite{Wang04}. In order to apply it
to the study of massive nuclear reactions we first make test if
this model is suitable for heavy nuclear fusion reactions.
Concerning the parameters for the energy density functional (see
\cite{Wang04}) a new set of parameter IQ2 is developed with which
both the capture cross sections and the process of quasi-fission
for heavy nuclear fusions can be described well, in addition to
the parameter set of IQ1 used in \cite{Wang04}. The study on the
process of the quasi-fission in heavy nuclear fusion reactions
will be given in other work. Both IQ2 and IQ1 can describe the
fusion reaction of light and intermediate nuclei well. All the
calculation results given in this work are obtained with IQ2,
which is given in table 1. Fig.1 shows the capture cross sections
of the $^{16}$O+$^{208}$Pb and $^{48}$Ca+$^{208}$Pb. One can see
that the capture cross sections obtained from the ImQMD model are
in good agreement with the experimental data \cite{Mor99,Zag01} at
the energies near and above Coulomb barrier. Another relevant test
to our present study is the charge distribution of products in
reactions between two heavy nuclei. As an example in Fig.2 we show
the charge distribution of products in the central collisions of
the $^{197}$Au+$^{197}$Au at 35 $AMeV$ and compared with
experimental data\cite{Dse98}. The agreement is quite satisfied.
We have also calculated the charge distribution for reactions of
light and intermediate mass systems. The nice agreement is also
obtained, which will shown in other publications.

Now let us apply the ImQMD model to strongly damped reactions of
$^{244}$Pu+$^{244}$Pu, $^{238}$U+$^{238}$U and
$^{197}$Au+$^{197}$Au at energy range of $E_{c.m.}$=800 $MeV$-2000
$MeV$. The impact parameters are taken to be 1 $fm$ and 3 $fm$.
The simulation events are taken to be 500 for each energy point
and impact parameter.  The initial nuclei of projectile and target
are prepared by the same procedure as in ref.
\cite{Wang02,Wang04}. Since in this work we mainly concern the
production of the superheavy fragments, for saving CPU time (this
kind calculation is very time consuming) the simulation procedure
is carried out as follows: In each event, the simulation is
continued until $t$=6000 $fm/c$ if there exists a superheavy
fragment . As soon as it is found that there no superheavy
fragment exists then the simulation is terminated. In this way we
can save a lot of CPU time.

Fig. 3 shows the energy dependence of the probability of producing
superheavy fragments for three reactions of $^{244}$Pu+$^{244}$Pu,
$^{238}$U+$^{238}$U and $^{197}$Au+$^{197}$Au at impact parameter
$b$=1 $fm$. It shows us that among these three reactions, the
yield of superheavy fragments produced in $^{244}$Pu+$^{244}$Pu is
the highest, and that produced in  $^{238}$U+$^{238}$U is only
half of the Pu+Pu's yield. For the $^{197}$Au+$^{197}$Au, we do
not find any reaction event which forms a product with Z $\geq$
114 in the present calculations. The very pronounced feature of
the figure is that it is narrowly peaked in the energy dependence
of the production probability of SHFs and the location of the peak
is at about $E_{c.m.}$ = 1000 $MeV$ for the $^{244}$Pu+$^{244}$Pu
and at $E_{c.m.}$ = 950 $MeV$ for the $^{238}$U+$^{238}$U.
Although the precise location of peak energy may not be very
definite in this primary calculation, such behavior of the energy
dependence of the probability of superheavy fragments with $Z
\geq$ 114 should be correct. The narrow peak means that it is
crucial to select the correct incident energy in order to search
superheavy elements experimentally by using the approach of the
strongly damped massive reactions. We notice that the energies
used in the experiments done in the 70s and 80s
\cite{Hil77,Gag80,Sch78} are lower than the peak energy for the
reaction of $^{238}U+^{238}U$. The production probability of SHFs
corresponding to the energies used in \cite{Hil77,Gag80,Sch78} is
much lower than that at the peak energy. The results about the
incident energy dependence of the probability of producing SHFs at
impact parameter $b$=3 $fm$ are quite similar with at $b$=1 $fm$.
For surveying the proton and neutron numbers in SHF obtained we
draw the contour plot of mass and charge distributions of SHFs
with $Z \geq$ 114 at the time $t$=6000 $fm/c$ for the reaction of
$^{244}$Pu+$^{244}$Pu in the inserted figure of Fig.3. For
comparison, the experimental data of isotopes of $^{288}$114,
$^{287,288}$115 and $^{292}$116 \cite{Oga04} are also given in the
figure by the black points. One can see from the figure that quite
a few SHFs in the reaction of Pu+Pu are very neutron rich and the
corresponding neutron-to-proton ratio is much higher than that
obtained experimentally. This character is very useful for
approaching to the center of  superheavy "island".

Now let us discuss the decay mechanism of the composite system of
projectile and target.  Fig.4 shows the time evolution of the
number of SHFs for the reaction of $^{244}$Pu+$^{244}$Pu at
$E_{c.m.}$=1000 $MeV$ and the $^{238}$U+$^{238}$U at $E_{c.m.}$=
950 $MeV$ with impact parameters $b$=1, 3 $fm$.  The number of
SHFs is obtained within 500 events for each impact parameter. From
Fig.4 two stages of the decay process of the composite systems can
be distinguished by very different decreasing slope, which implies
very different decay mechanism of the composite system. From 1000
$fm/c$ to 1500 $fm/c$, the number of SHFs decreases quickly with
time increasing. During this stage, the composite system firstly
breaks up into two pieces, which we call the first decay. We have
counted the number of the existing composite systems at different
time. At $t$=1000 $fm/c$, more than 60 percent of events are still
in the stage of two reaction partners sticking together, then at
1200 $fm/c$, about 10-15 percent of events remain in this stage,
and at 1500 $fm/c$ only few remains in this stage, i.e. almost all
composite systems break up into two pieces. In the most of cases,
the composite system breaks up into two pieces with size close to
the initial nuclei. In a few cases it breaks up into two pieces
with one heavier fragment and another smaller fragment, and in
this case there is possibility to produce one SHF and its partner
with $Z\sim 70$. Then some of SHFs further break up into two
pieces quickly within several tens and hundreds fm/c and some of
SHFs survive followed by the slow decreasing stage. In the second
stage, the number of SHFs decreases slowly with time. In this
stage, the number of SHFs is reduced through emitting light
charged particles, protons accompanying with neutron emission, and
also still breaking into two pieces process. The slow reduction of
the number of SHFs in the second stage seems to be benefit to the
survival of SHFs.

In Fig.5 we show the distributions of both a)the binding energies
and b)the $R_{z}/R_{\rho}$ of SHFs produced in the reaction of
$^{244}$Pu+$^{244}$Pu at $E_{c.m.}$=1000 $MeV$ and 950 $MeV$, and
$b$=1 $fm$. The $R_{z}$ is the long axis and $R_{\rho}$ is the
short axis of SHF. The figure was drawn as the count number in
1000 reaction events vs a)the binding energies and b)the values of
$R_{z}/R_{\rho}$. From Fig.5a) one sees that the binding energies
of SHFs are broadly distributed. In the large binding energy side,
the binding energy reaches about 7 $MeV/nucleon$, which is not far
from the value of the predicted binding energy of the ground state
of corresponding superheavy elements. The feature of broad
distribution of binding energies of SHFs tailing to large binding
energy is favorable to have larger surviving probability of SHF.
From Fig.5b one sees that the SHFs are strongly deformed. In the
most of cases, they are at about super-deformation or even
hyper-deformation. For those SHFs with super-deformed shape it is
found that there are some bubbles in the density distribution
(bubble-like). However, there also exist some exotic forms among
the produced SHFs with $R_{z}/R_{\rho}\geq$ 4. The shape of these
SHF is band-like. It is very surprising that the shape of SHFs has
such exotic form. Such exotic forms of SHFs may be attributed to
the huge electric charge. Associating the recent structure studies
of superheavy nuclei within the RMF and HFB theory
\cite{Ben01,Xu04,Ren02} in which very large deformed isomeric
states were predicted, the exotic form of SHFs seems to be
understandable. However, the subject of exotic (bubble,band-like)
configurations in super-heavy elements in which the interplay
between Coulomb interaction and nuclear interaction becomes very
important needs to be further studied.

 In summary, within the microscopically dynamical
description of the $^{244}$Pu+$^{244}$Pu, $^{238}$U+$^{238}$U and
$^{197}$Au+$^{197}$Au reactions we  explore the dynamic process of
the strong damped reaction and find that the production
probability of superheavy fragments with $Z \geq$ 114 in the
$^{244}$Pu+$^{244}$Pu reaction is much higher than that in the
$^{238}$U+$^{238}$U reaction, and no product of SHF has been found
for the $^{197}$Au+$^{197}$Au reaction in the present study. The
narrowly peaked energy dependence of the production probability of
SHFs shows that the suitable selection of the incident energy is
very important for searching for super-heavy elements by means of
strongly damped massive nuclear reactions. The dynamical study
shows that there are two different kind mechanisms in the decay of
the composite system. The first stage is a fast process composed
by breaking the composite system into two pieces and further
breaking the heavy fragments into two small pieces. The second
stage is a slow process by emitting light charged particles and
nucleons as well as further breaking of SHFs. This study also
explores that the binding energies of SHFs are distributed
broadly. Its tail at large binding energy side is not far from the
predicted binding energy of the corresponding SHE and therefore is
favorable to producing superheavy elements. The shape of SHFs is
also studied and turns out to be strongly deformed. It seems to us
that the study on the structure and the fission barrier for such
exotic shape due to extremely strong Coulomb effect is urgently
required in order to learn if the stabilized superheavy nuclei can
be eventually reached or not. This study is still in progress.

\begin{center}
{\bf ACKNOWLEDGMENTS}
\end{center}
This paper is supported by the China Postdoctoral Science
Foundation, the National Natural Science Foundation of China under
Grant Nos. 10175093,10175089,10235030,10235020,10347142, Major
State Basic Research Development Program under Contract No.
G20000774 and CAS-grant KJCX2-SW-N02.

\newpage

\begin{center}
{\small {\bf CAPTIONS} }
\end{center}

\begin{description}

\item[{\tt Fig.1}] The capture cross sections of  the reactions of
$^{16}$O+$^{208}$Pb and $^{48}$Ca+$^{208}$Pb as a function of
incident energies.

\item[{\tt Fig.2}]  The charge distribution of the central
collisions of $^{197}$Au+$^{197}$Au  at 35 $AMeV$. The simulation
are ended at 6000 $fm/c$. The open and solid circles denote the
calculated results and experimental data \cite{Dse98},
respectively.

\item[{\tt Fig.3}] The incident energy dependence of the
production probability of superheavy fragments with $Z \geq$ 114
in reactions of $^{244}$Pu+$^{244}$Pu and $^{238}$U+$^{238}$U at
impact parameter $b$=1 $fm$. The inserted figure is the contour
plot of mass and charge distributions of the products with Z
$\geq$ 114 at the time $t$=6000 $fm/c$, in which the solid circles
denote the experimental data of isotopes of $^{288}$114,
$^{287,288}$115 and $^{292}$116 \cite{Oga04}.

\item[{\tt Fig.4}] The time evolution of the number of fragments
with $Z \geq$114 including the heavy residues of composite systems
for the reactions of $^{244}$Pu+$^{244}$Pu at E$_{c.m.}$=1000
$MeV$ and $b$=1 and 3 $fm$, and for the reactions of
$^{238}$U+$^{238}$U at E$_{c.m.}$=950 $MeV$ and $b$=1 and 3 $fm$
from time $t$=1000 to 6000 $fm/c$.

\item[{\tt Fig.5}] The distributions of a)the binding energies and
       b)the $R_{z}/R{\rho}$ of SHFs produced in the reaction of
       $^{244}$Pu+$^{244}$Pu at $E_{c.m.}$=1000 $MeV$ and 950 $MeV$, and $b$=1 $fm$. The
       count number is obtained for 1000 reaction events.

\item[{\tt Table.1}] The model parameter IQ2.
\end{description}
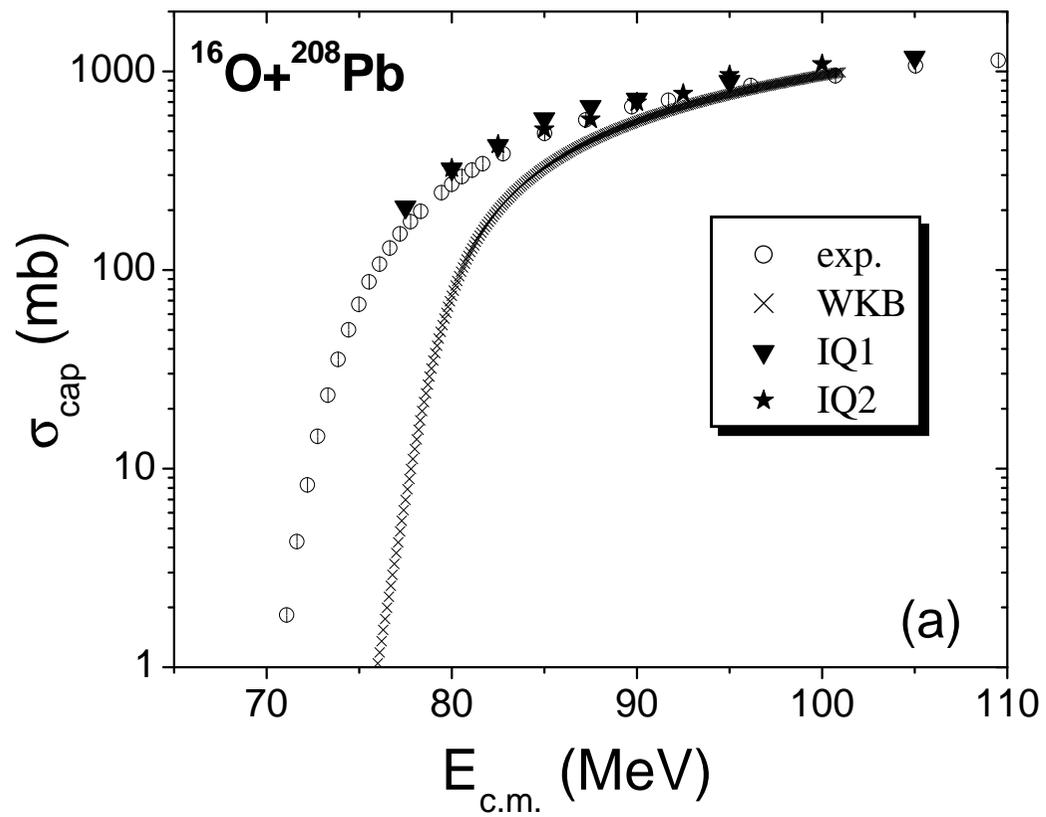
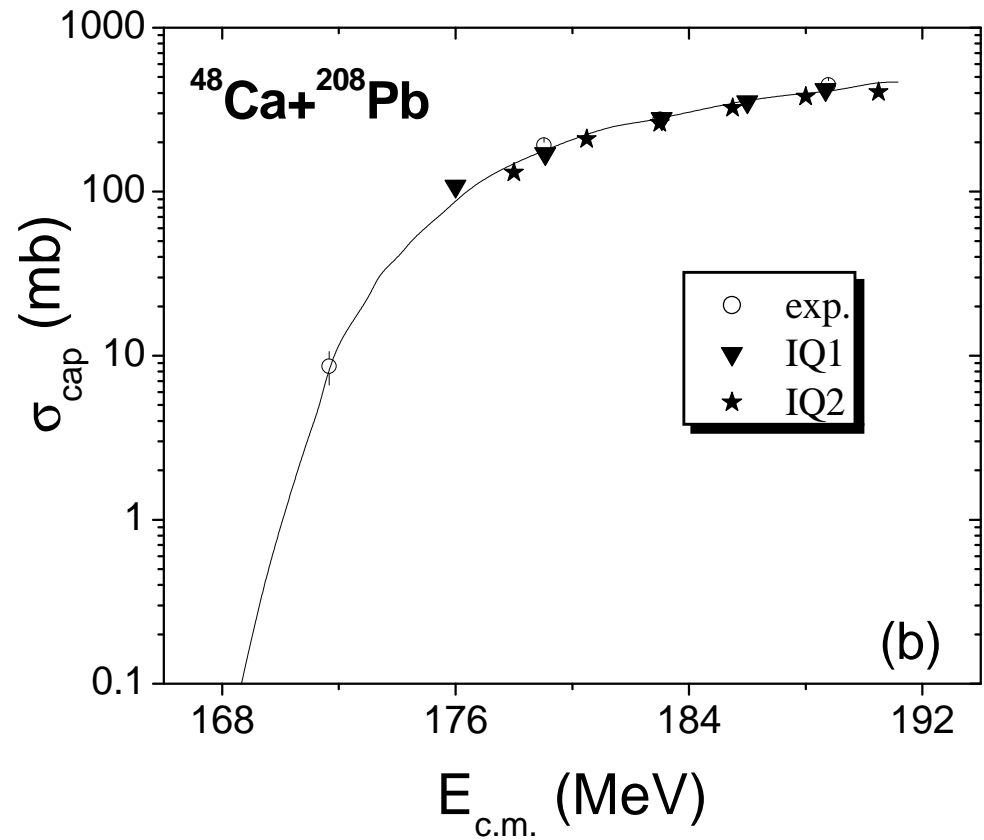
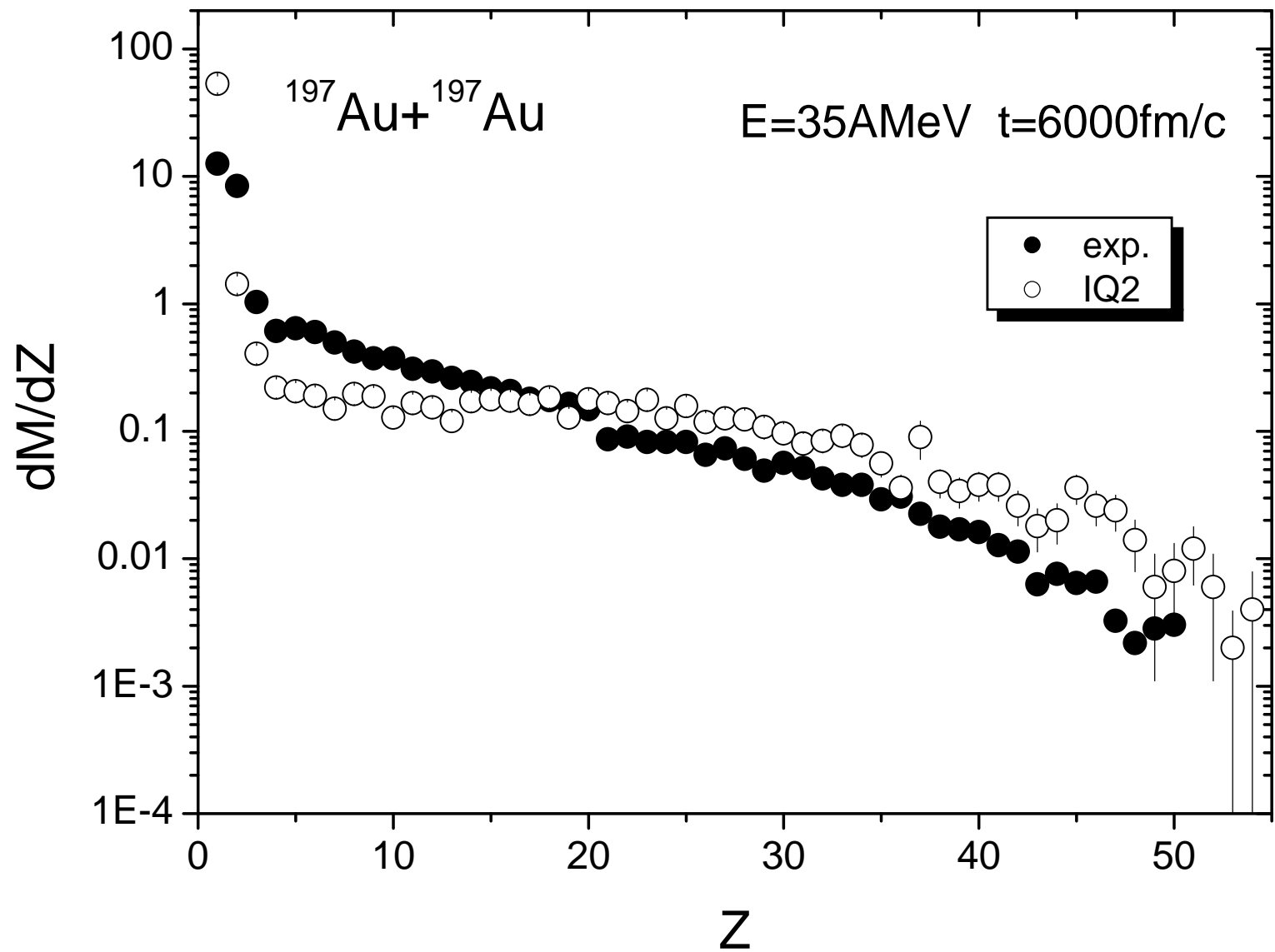
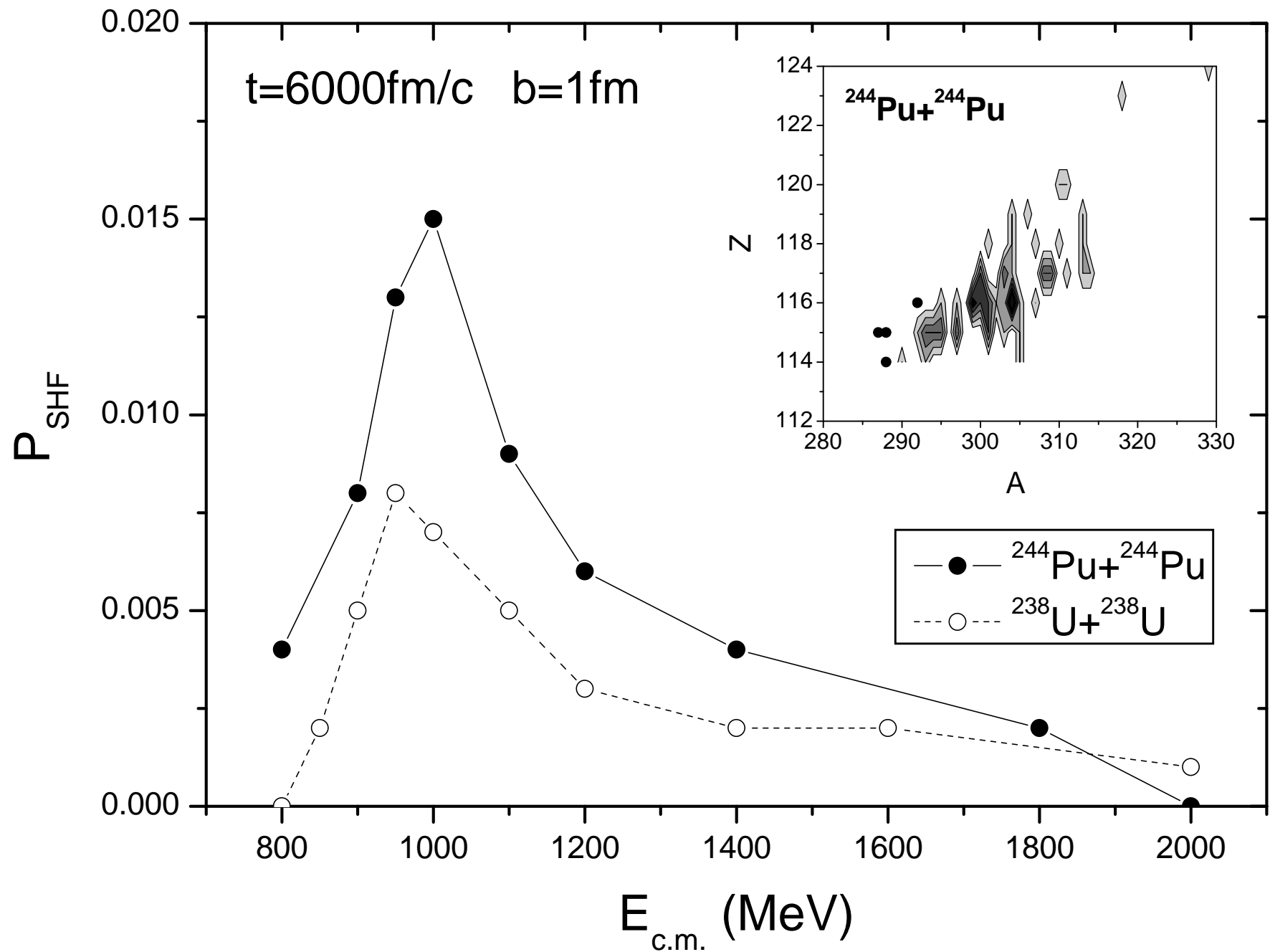
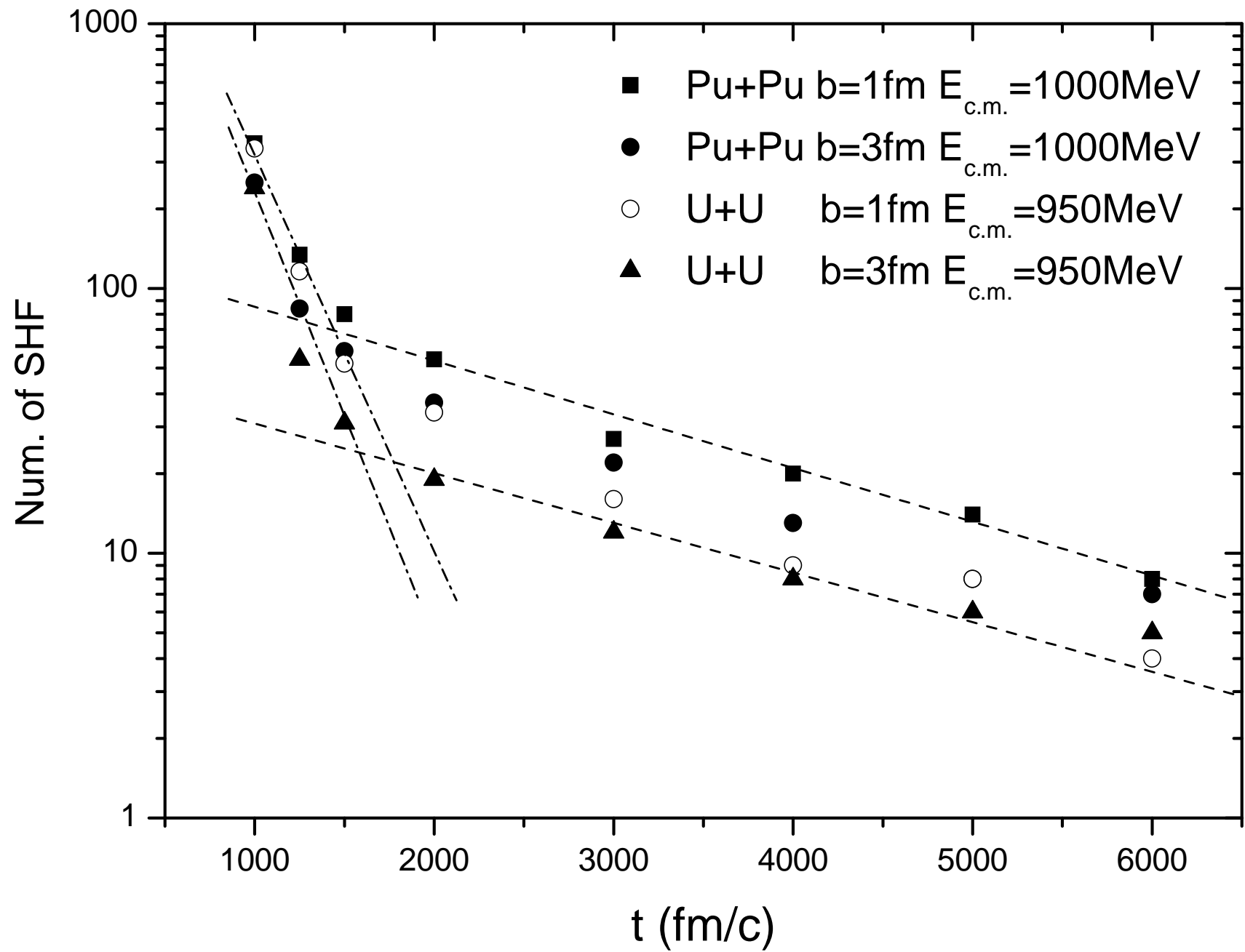
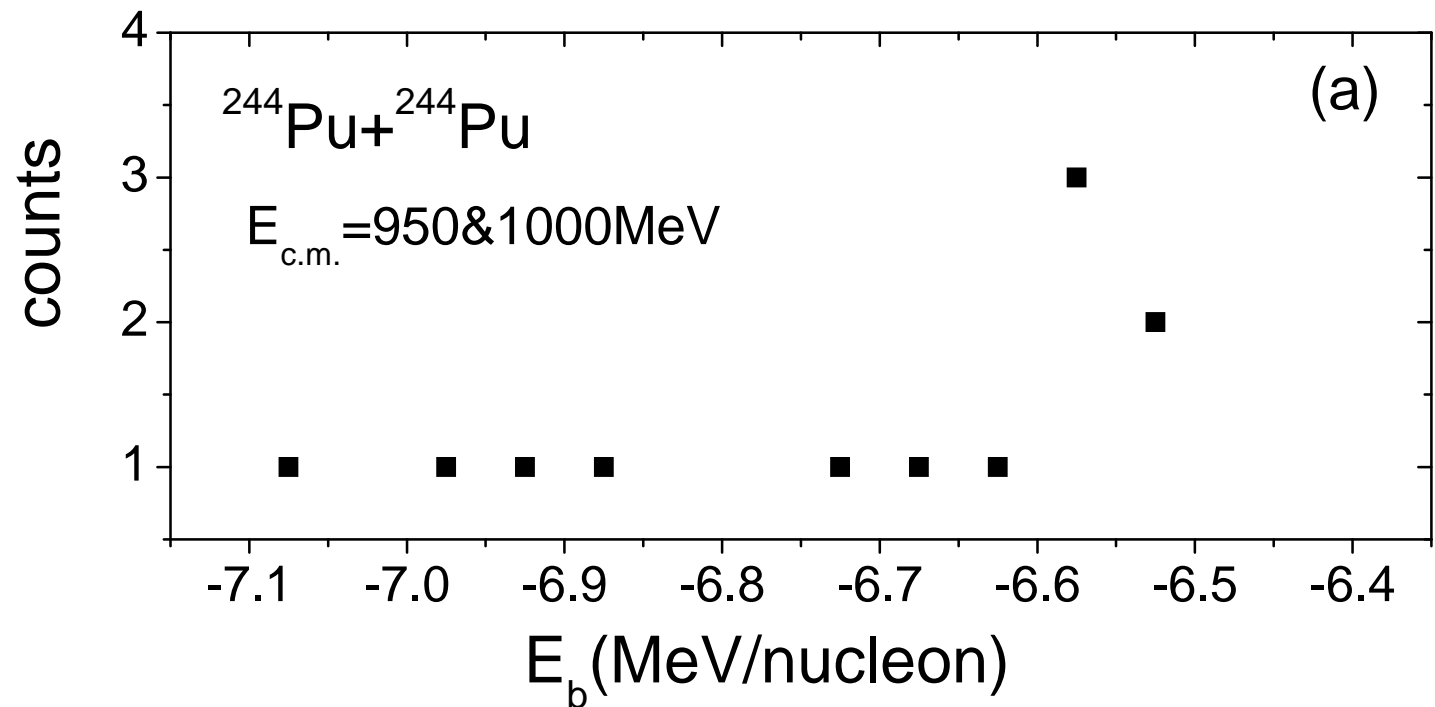
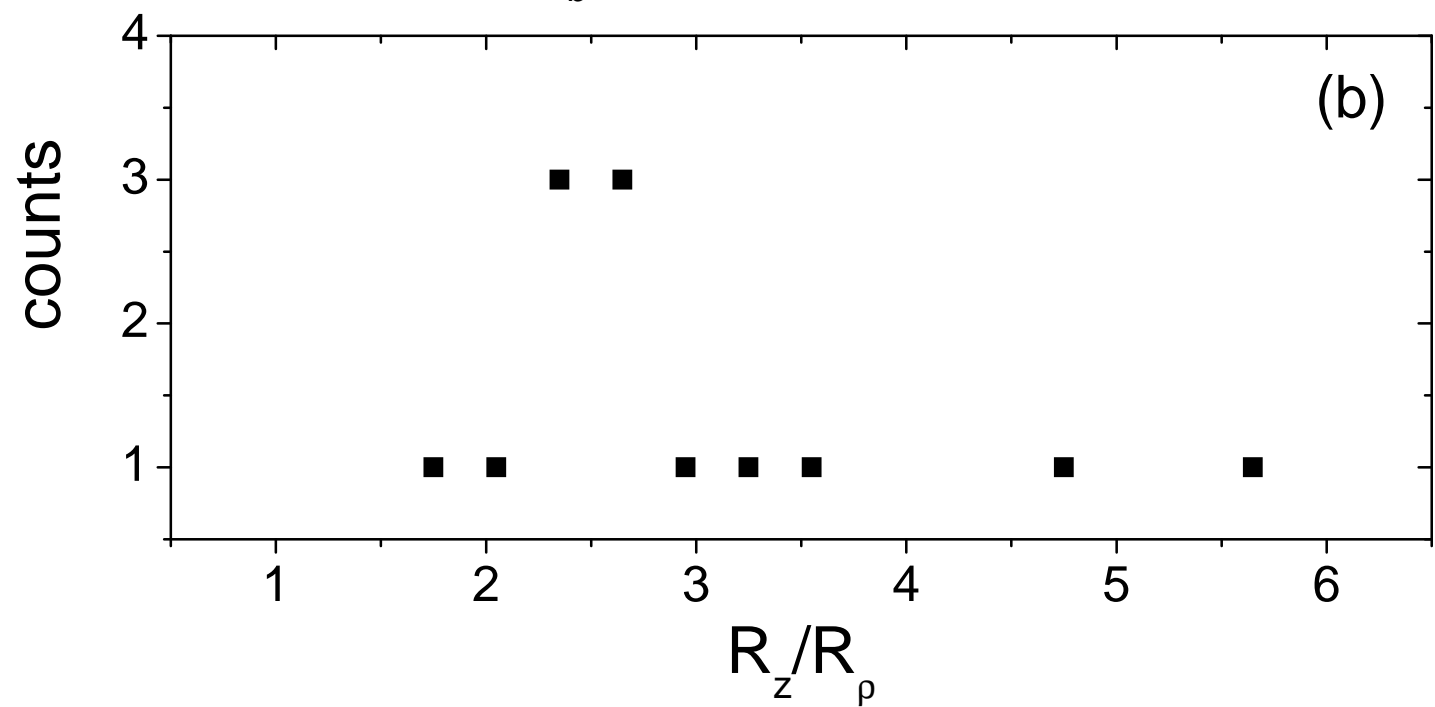
\end{document}